\documentclass[%
 reprint,
 amsmath,amssymb,
 aps,
]{revtex4-2}

\usepackage{graphicx}% Include figure files
\usepackage{dcolumn}% Align table columns on decimal point
\usepackage{bm}% bold math
\usepackage[squaren,Gray,cdot]{SIunits}
\usepackage{hyperref}% add hypertext capabilities
\usepackage[version=4]{mhchem}
\begin{document}

\preprint{APS/123-QED}

\title{Cavitation cloud formation and surface damage of a model stone in a high-intensity focused ultrasound field}

\author{Luc Biasiori-Poulanges}
    %\email[Correspondence email address: ]{lbiasiori@ethz.ch}
    \affiliation{ETH Zurich, Department of Mechanical and Process Engineering, Institute of Fluid Dynamics, Sonneggstrasse 3, Zurich 8092, Switzerland}

\author{Bratislav Luki\'c}
    %\email[Correspondence email address: ]{bratislav.lukic@esrf.fr}
    \affiliation{European Synchrotron Radiation Facility, CS 40220, Grenoble F-38043, France}
    
\author{Outi Supponen}
    \email[Correspondence email address: ]{outis@ethz.ch}
    \affiliation{ETH Zurich, Department of Mechanical and Process Engineering, Institute of Fluid Dynamics, Sonneggstrasse 3, Zurich 8092, Switzerland}

\date{\today}

\begin{abstract}
This work investigates the fundamental role of cavitation bubble clouds in stone comminution by focused ultrasound. The fragmentation of stones by ultrasound has applications in medical lithotripsy for the comminution of kidney stones or gall stones, where their fragmentation is widely assumed to result from the high acoustic wave energy. However, high-intensity ultrasound can generate cavitation which is known to contribute to erosion as well and to cause damage away from the target, although the exact contribution of cavitation remains currently unclear. Based on \textit{in-situ} experimental observations, \textit{post-mortem} microtomography and acoustic simulations, the present work sheds light on the fundamental role of cavitation bubbles in the stone surface fragmentation by correlating the detected damages to the observed bubble activity. Our results show that not all clouds erode the stone, but only those located in preferential nucleation sites whose locations are herein examined. Furthermore, quantitative characterizations of the bubble clouds and their trajectories within the ultrasonic field are discussed. These include experiments with and without the presence of a model stone in the acoustic path length. Finally, the optimal stone-to-source distance maximizing the cavitation-induced surface damage area has been determined. Assuming the pressure magnitude within the focal region to exceed the cavitation pressure threshold, this location does not correspond to the acoustic focus, where the pressure is maximal, but rather to the region where the acoustic beam and thereby the acoustic cavitation activity near the stone surface is the widest.
\end{abstract}

\maketitle

%\tableofcontents

\section{\label{sec:introduction}Introduction}

Lithotripsy is a non-invasive clinical therapy for nephro- and ureterolithiasis~\cite{lingeman2009shock,leighton2010lithotripsy}, which consists in the fragmentation of renal and ureteric calculi when exposed to a strong external pressure, such as pulsed shocks as used in extracorporeal shock wave lithotripsy (ESWL)~\cite{miller2012overview,brennen2015cavitation,bader2019whom,pishchalnikov2003cavitation}. The ability of such high-amplitude pulses to fragment stones is believed to be attributed to both the high wave energy and the generation of cavitation bubbles~\cite{arora2005cavitation,loske2010role,lauterborn2015acoustic,pishchalnikov2019high}. However, given that the cavitation bubbles are induced by the shock wave, dissociating the contribution of the wave energy itself from that of the induced vapor bubbles is particularly complicated. In the absence of direct evidence, no consensus has yet been found and the exact physical mechanism resulting in the stone comminution remains unclear. Based on the large amount of work reporting on cavitation-induced damage for decades~\cite{plesset1955mechanism,plesset1977bubble,caupin2006cavitation,blake1987cavitation,brennen2015cavitation,soyama2020cavitating}, the role of cavitation bubbles in lithotripsy is supposed as follows. Upon exposure to the negative phase of the pulsed shock wave (\textit{i.e.}, the tensile component), cavitation bubbles are formed near the stone surface. These bubbles then scatter and absorb the energy of the subsequent pulses until they violently collapse. The collapse of individual bubbles initiates a high-speed jet, which develops towards neighbouring boundaries, that is, in the direction of the stone. The collapse of cavitation bubbles are known to produce shock waves and jets~\cite{tomita1986mechanisms}, which both exhibit a high cavitation damage potential and are suspected to erode the calculus during lithotripsy treatments, potentially contributing to the complete ablation of the stone~\cite{zhong2013shock}. This belief is further supported by past reports that the damage potential of a lithotripter shock wave on an artificial kidney stone is reduced in degassed water that contains fewer cavitation nuclei compared to a gas-saturated medium~\cite{loske2010role}. Furthermore, another study demonstrated the role of cavitation by applying a two-frequency ultrasound waveform specifically designed to amplify the ultrasound-induced cavitation bubble cloud collapse, which was found to enhance the stone erosion~\cite{ikeda2006cloud}.
Laser-induced cavitation relevant for intracorporeal laser lithotripsy~\cite{ho2021role,chen2022cavitation}, and especially the shock waves emitted from the asymmetric collapse of the large cavitation bubble~\cite{xiang2023dissimilar}, has been demonstrated to represent the major damage mechanism for the model stone instead of photothermal ablation from the laser.
Despite these recent advancements and knowledge on the various relevant cavitation-induced mechanisms, which have been extensively reviewed in the context of ESWL in ref.~\cite{zhong2013shock}, the exact damage contribution of bubble clouds typical for focused sound waves such as shock waves or HIFU remains challenging to quantify.

Extracorporeal shock wave lithotripsy has been shown to result in strong side effects~\cite{lingeman1989role} such as the erosion of the healthy neighboring tissues or hemorrhage~\cite{stoller1989severe,silberstein2008shock}. In the attempt to limit these injurious adverse effects, by reducing the source energy, high-intensity focused ultrasound-based lithotripsy has recently been proposed as an alternative method to ESWL~\cite{maxwell2015fragmentation,duryea2011vitro,duryea2013controlled,ikeda2016focused,yoshizawa2009high,ikeda2006cloud,connors2014comparison}. It can damage the stone with long bursts of lower-amplitude sound waves compared to ESWL. However, it has been shown that the size of the cavitation bubble cloud generated is in the order of the incident pressure wavelength~\cite{maeda2019bubble}. Consequently, the cloud strongly scatters the acoustic beam which results in the energy shielding of the stone. This reduces the stone exposure to the incident radiation and so the stone comminution efficiency. 

This work investigates the fundamental role of cavitation bubble clouds in the stone fragmentation through laboratory-scale experiments using a focusing ultrasound transducer and acoustic numerical simulations. We first describe the bubble cloud formation, properties and trajectories in the absence of a stone in the acoustic path. We then position a model stone in the focal region of the transducer and characterize its effect on the cloud propagation. Observing surface fragmentation and damages at the proximal region of the stone, we eventually correlate the reported damages to the bubble activity, and therefore directly evidence the role of bubble cloud in the stone surface fragmentation. Finally, we discuss the optimal source-to-stone distance to maximize the surface damage.

% **********************************************************
% ******************       Methods        ******************
% **********************************************************

\section{\label{sec:methods}Methods}

The role of cavitation bubbles on the stone fragmentation is addressed by means of experimental observations as well as time-domain acoustic simulations using a $k$-space pseudospectral method.

In this work, results are discussed based on dimensionless parameters. The axial and radial coordinates $z$ and $r$ of the pressure field are normalized by the theoretical acoustic focus $\gamma_f$ and the full width at half maximum of the ellipsoidal focal region $\ell_1$, respectively. Unless otherwise specified, the non-dimensionalization of the pressure magnitude $p$ is done by using the maximum pressure, $\max\{p\}$, which is the pressure at the acoustic focus. The resulting change in variables is
\begin{equation}
    z^{*} = \frac{z}{\gamma_f}, \quad
    r^{*} = \frac{r}{\ell_1}, \quad
    p^{*} = \frac{p}{\max\{p\}},
    \label{eq:conventions}
\end{equation}
where $(\cdot)^{*}$ denotes a non-dimensional quantity. Approximations in the analytical derivation of the  acoustic focus yields $\gamma_f \approx \beta[1+{2\beta^2}/({1+2\beta})]^{-1}$~\cite{huang2009simple}, where $\beta=f\lambda/a$ with $\lambda$ being the source wavelength, and $a$ and $f$ the radius and the geometrical focus of the transducer, respectively. Details on the analytical derivations used to calculate the theoretical axial coordinate of the physical focal distance $\gamma_f$, which corresponds to the axial coordinate of the maximum pressure, are given in Appendix~\ref{app:a}.

% ******************  Experimental set-up  *****************

\subsection{\label{subsec:exp_setup}Experimental set-up}

\begin{figure}[htbp]
    \centering
    \includegraphics[width=.65\columnwidth]{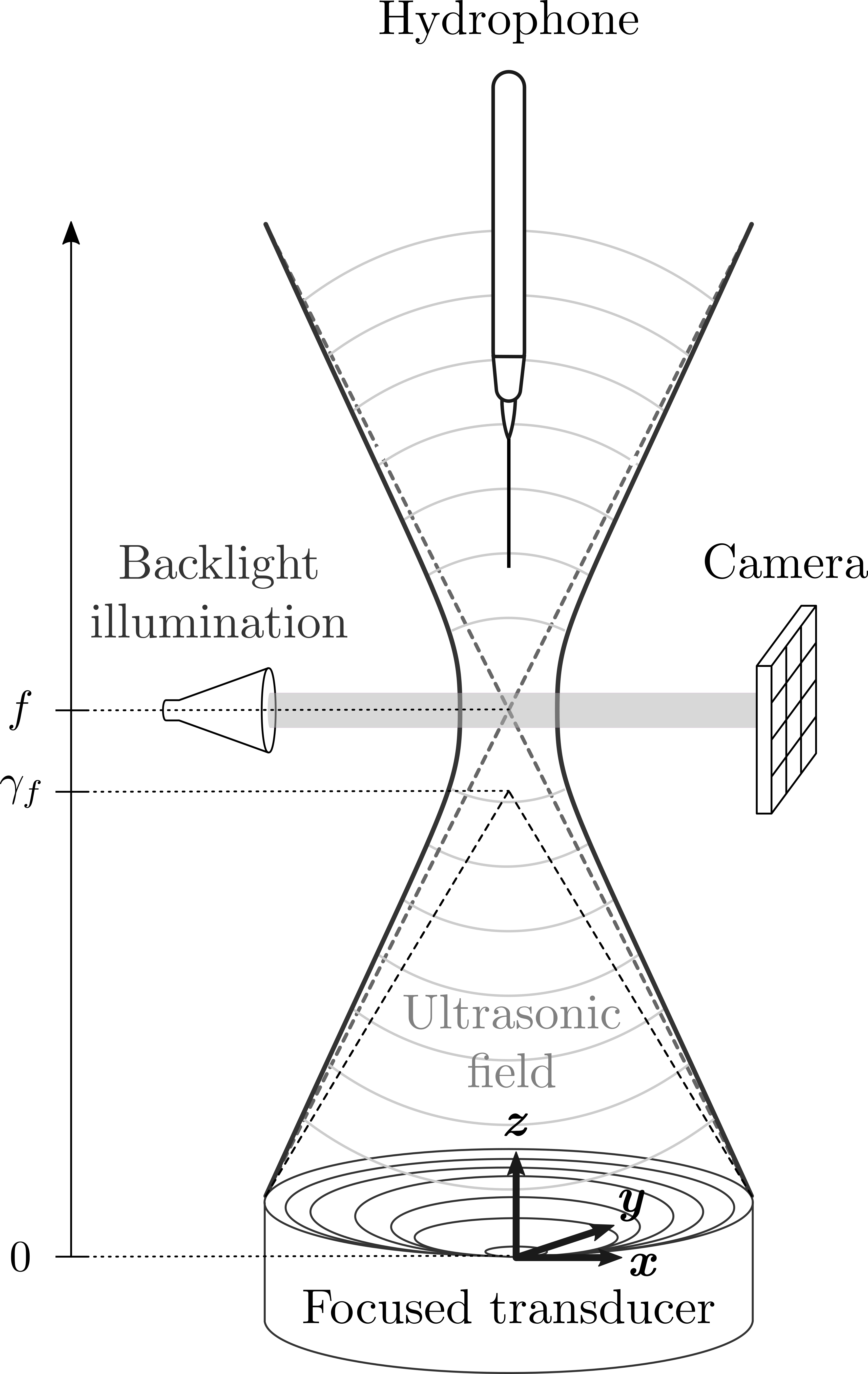} %reprint
    \caption{Schematics of the experimental set-up}
    \label{fig:1}
\end{figure}

A schematic of the experimental set-up is shown in Fig.~\ref{fig:1}

The high-intensity focused ultrasound (HIFU) field is generated using a single element curved bowl-shaped piezo-ceramic transducer (Precision Acoustics). The nominal central frequency and the acoustic path length are $250~\kilo\hertz$ and $67~\milli\meter$, respectively. 

The drive signal is generated with an arbitrary waveform generator (DG1032Z, Rigol) coupled to $200$-$\watt$ radio frequency power amplifier (1020L, Electronics \& Innovation). The transducer is operated in burst mode where 500 cycles of a sinusoidal signal are applied over a 20-$\milli\second$ period, which corresponds to a $10$-$\%$ duty cycle. The input voltage is $180~\volt_{\mathrm{pp}}$. The transducer is immersed and centered at the bottom of a square base test chamber filled with tap water, which has been here used to increase the cavitation probability. The acoustic radiation is directed from bottom to top. The rectangular test chamber is $300~\milli\meter\times300~\milli\meter\times350~\milli\meter$ and consists in transparent glass walls that allow for optical access. The ultrasound field is measured using a 1-$\milli\meter$ piezoelectric needle hydrophone (Precision Acoustics) mounted on a 3-direction micro-translation stage, with the electrical signal recorded by a fast digital oscilloscope (DS1104Z Plus, Rigol). Note that to avoid generating cavitation-induced damages on the hydrophone while measuring the pressure field, the transducer has been operated at half the maximum intensity (i.e., $90~\volt_{\mathrm{pp}}$, which corresponds to a peak pressure $\max\{p\}\approx0.2~\mega\pascal$.). Note that, in the context of such lab-scale experiments, the HIFU transducer generates relatively weak acoustic waves compared to clinically relevant transducers. However, the acoustic amplitudes suffice to nucleate cavitation in the bulk of the liquid.

Model kidney stones are made with BegoStone, which is a superhard plaster composed of Calcium Sulfate Hemihydrate (Ca$_{2}$H$_{2}$O$_{9}$S$_{2}$), mixed with distilled water to a $15:6$ weight ratio. The mixture is magnetically stirred during $60~\second$, before being degassed using a vacuum chamber equipped with a vacuum pump. The mixture is eventually poured into 20-\milli\meter-diameter hemispherical silicone moulds where the samples are allowed to cure for 12-24 hours at room temperature. The spherical face of the phantom kidney stones are located at the focal point of the HIFU field and exposed to the incident acoustic beam.

The cavitation bubble clouds are imaged with shadowgraphy recorded with a high-speed camera (Photron Fastcam Nova S12) equipped with a $100~\milli\meter~f/2.8~2\times$ macro lens (Laowa). The exposure time is $210~\nano\second$. The sampling frequency, not constant from one experiment to another, is indicated in the caption of the figures. The backlight illumination is given by a broadband halogen fiber optic illuminator which provides a continuous collimated $150$-$\watt$ light operating over the 400-1600~$\nano\meter$ range. 

% ****************  Numerical simulations ***************

\subsection{\label{subsec:num_simu}Numerical simulations}

Acoustic simulations are carried out using the k-Wave MATLAB toolbox, which uses a $k$-space pseudospectral time domain method to solve the discretized acoustic wave equations in a Cartesian coordinate system. It accounts for a frequency power law absorption of the form $\alpha_{0}\omega^{y}$, where $\omega$ is the temporal frequency and $\alpha_{0}$ is the attenuation coefficient in water that equals to $0.0022~\mathrm{dB}\per\mega\hertz^{y}\per\centi\meter$, with $y=2$. The problem is treated in a two-dimensional formulation with axisymmetric conditions. The source is modeled as the injection of mass within the computational domain, where boundary conditions account for the tank walls, the hemispherical stone (if present) and the concave transducer. This was performed by specifying time-varying source pressure for each grid point meshing the source geometry. 

\begin{figure}[htbp]
    \centering
    \includegraphics[width=.75\columnwidth]{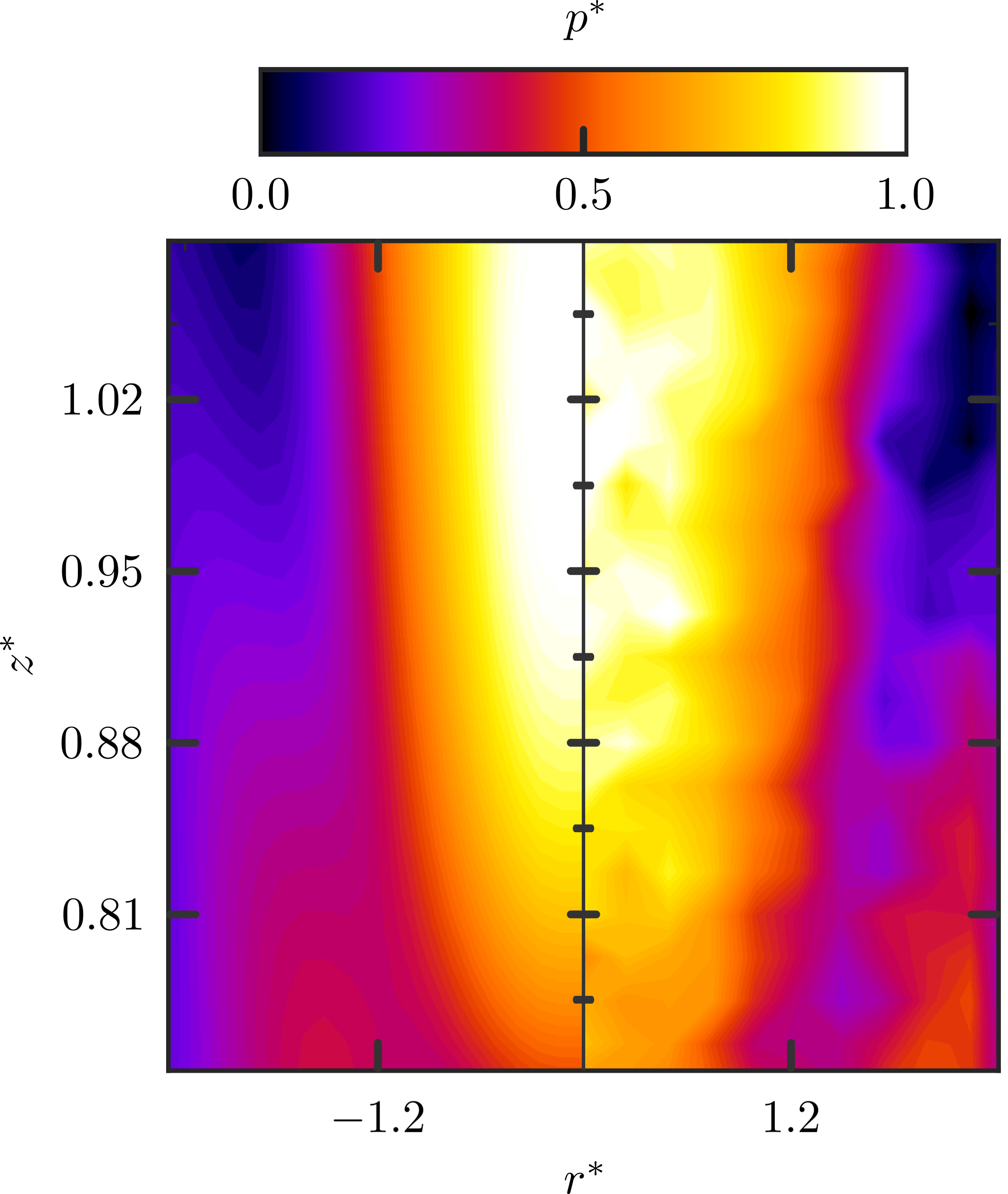} %reprint
    \caption{Comparison of the normalized pressure amplitude measured with the high-speed needle hydrophone (right) against the k-wave numerical simulation (left).}
    \label{fig:ExpVsNum}
\end{figure}

The computed pressure field is compared to the experimental measurement in Fig.~\ref{fig:ExpVsNum}. A very good agreement is found between experiments and numerics in both the pressure distribution and amplitude, and the acoustic beam shape near the focus. The geometry of the focal region, near the acoustic focus, is known to shape an elongated ellipsoid with semi-minor, $\ell_1$, and semi-major axis, $\ell_2$, given by $\ell_k=\alpha_k\lambda({f}/{2a})^k$, where $\ell_1$ is the full width at half maximum of the acoustic beam. Parameters $\alpha_1$ and $\alpha_2$ are fitting coefficients that we numerically found to be 1.26 and 8.24, respectively.

% *************  Influence of HIFU intensity ************

\subsection{\label{subsec:influence_intensity}Influence of the source intensity on the acoustic beam properties}

As previously mentioned, the pressure field has been experimentally characterized with the transducer running at half of its maximum intensity to prevent cavitation-induced damage on the hydrophone. To evaluate the change in the acoustic beam shape, it is instructive to evaluate the axial pressure by solving the O'Neil integral \cite{o1949theory}, which accounts for the transducer curvature and assumes a uniform distribution of the normal vibrational velocity over the transducer surface.

\begin{figure}[htbp]
    \centering
    \includegraphics[width=\columnwidth]{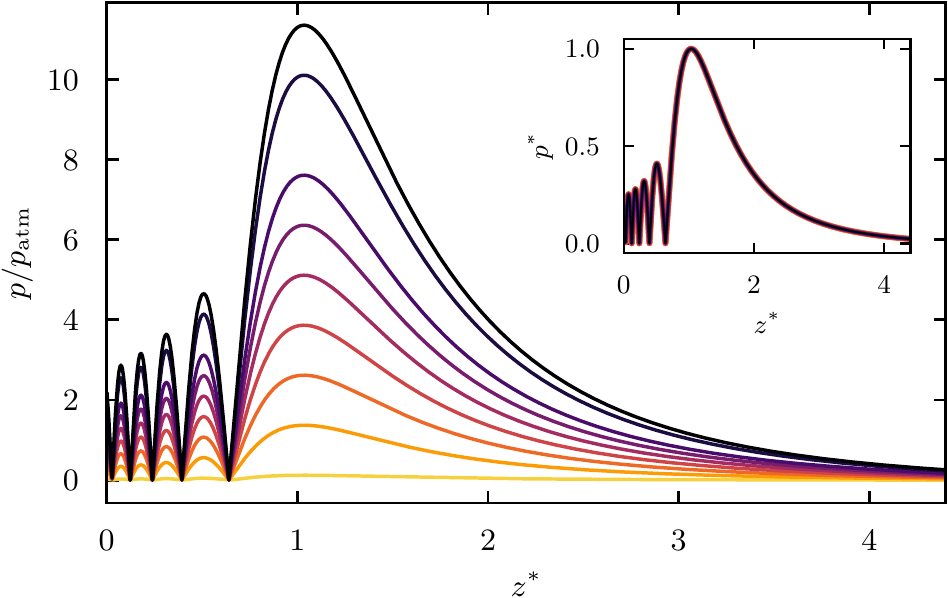} %reprint
    \caption{Axial pressure computed from the O'Neil integral for normal velocity ranging from 1 to 100~\milli\meter\per\second~by step of 10~\milli\meter\per\second. The pressure is normalized to atmospheric pressure $p_{\rm atm}=10^5$~Pa. The colormap goes from yellow to dark purple for increasing velocities. The inset plots the corresponding dimensionless axial pressure profiles as defined by Eq.~(\ref{eq:conventions}).}
    \label{fig:NormalVelocitySolution}
\end{figure}

\begin{figure}[htbp]
    \centering
    \includegraphics[width=\columnwidth]{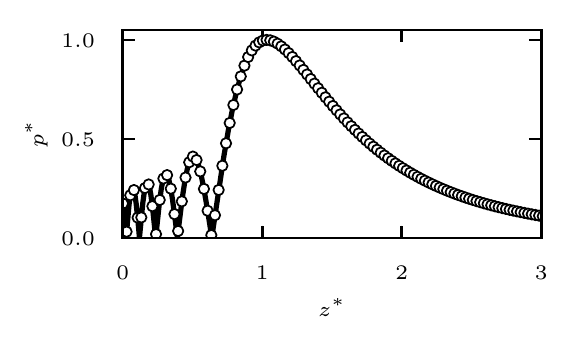} %reprint
    \caption{Validation of the numerical simulations using $k$-wave (circle markers) against the O'Neil integral solution (solid line).}
    \label{fig:ONeilSolution}
\end{figure}

Figure~\ref{fig:NormalVelocitySolution} plots the O'Neil integral solutions for normal velocities ranging from 1 to 100~\milli\meter\per\second~by step of 10~\milli\meter\per\second, in both dimensional and dimensionless forms. Not surprisingly, and expected for the pressure magnitude, it shows no change in the axial pressure profile. The location of the pressure lobes and the focal region remain unchanged. After validation of the numerical simulations against the O'Neil integral solution (Fig.~\ref{fig:ONeilSolution}), parametric simulations have also demonstrated that $\ell_1$ and $\ell_2$ are not affected by the initial transducer intensity. In light of these results, it is assumed that the geometry of the acoustic beam is independent of the intensity of the transducer, while the pressure is not. 

% **********************************************************
% ******************       Results        ******************
% **********************************************************

\section{\label{sec:results}Results and discussion}

In this work, we contribute to elucidating the role of the cavitation bubbles on the calculi fragmentation by correlating the location of the bubble clouds with the observation of the falling stone fragments. To collect the total cloud population in a single experiment which eventually results in the stone surface damage, we compiled time average shadowgraphs by using a minimum intensity projection algorithm (MinIP). The MinIP algorithm consist in selecting pixels of the lowest intensity from every frame $\mathfrak{f}$ throughout the time sequence to reconstruct a 2-D image. The intensity $I_r(i,j)$ of the reconstructed pixel located in $(i,j)$ is given by $I_r(i,j)=\min\{I(i,j)\}|_{\mathfrak{f}=1,2,...n}$, where $n$ is the total number of frames.

\begin{figure}[htbp]
    \centering
    \includegraphics[width=\columnwidth]{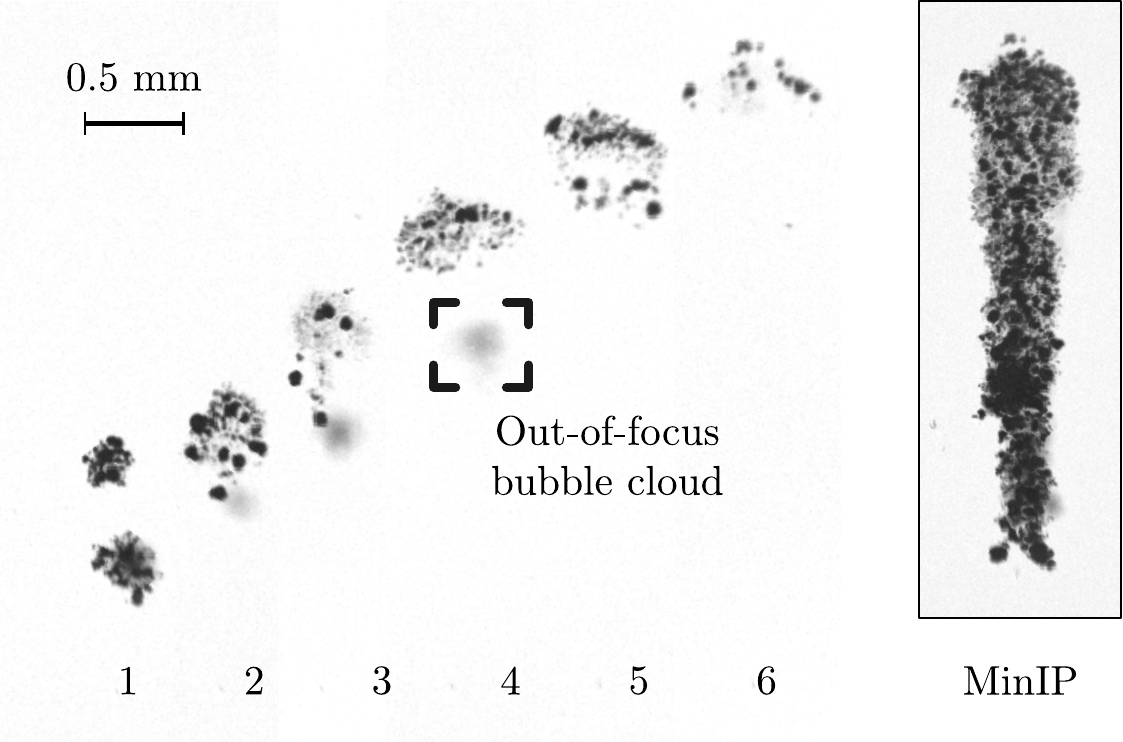} %reprint
    \caption{Time sequence of ultrasound-driven bubble clouds showing 6 frames out of 68, with a time interval of 333~\micro\second. The sampling frequency is 30~\kilo\hertz. The framed image on the right-hand side corresponds to the minimum intensity projection (MinIP) of the 68 frames.}
    \label{fig:minIP}
\end{figure}

Figure~\ref{fig:minIP} illustrates the use of the MinIP algorithm where 68 frames have been used to compile the reconstructed image of an upward-propagating bubble cloud initially formed from the merging of two subclouds. 

% *************  Bubble-cloud in free field ************

\subsection{Bubble cloud in a stone-free field}

\begin{figure}[htbp]
    \centering
    \includegraphics[width=\columnwidth]{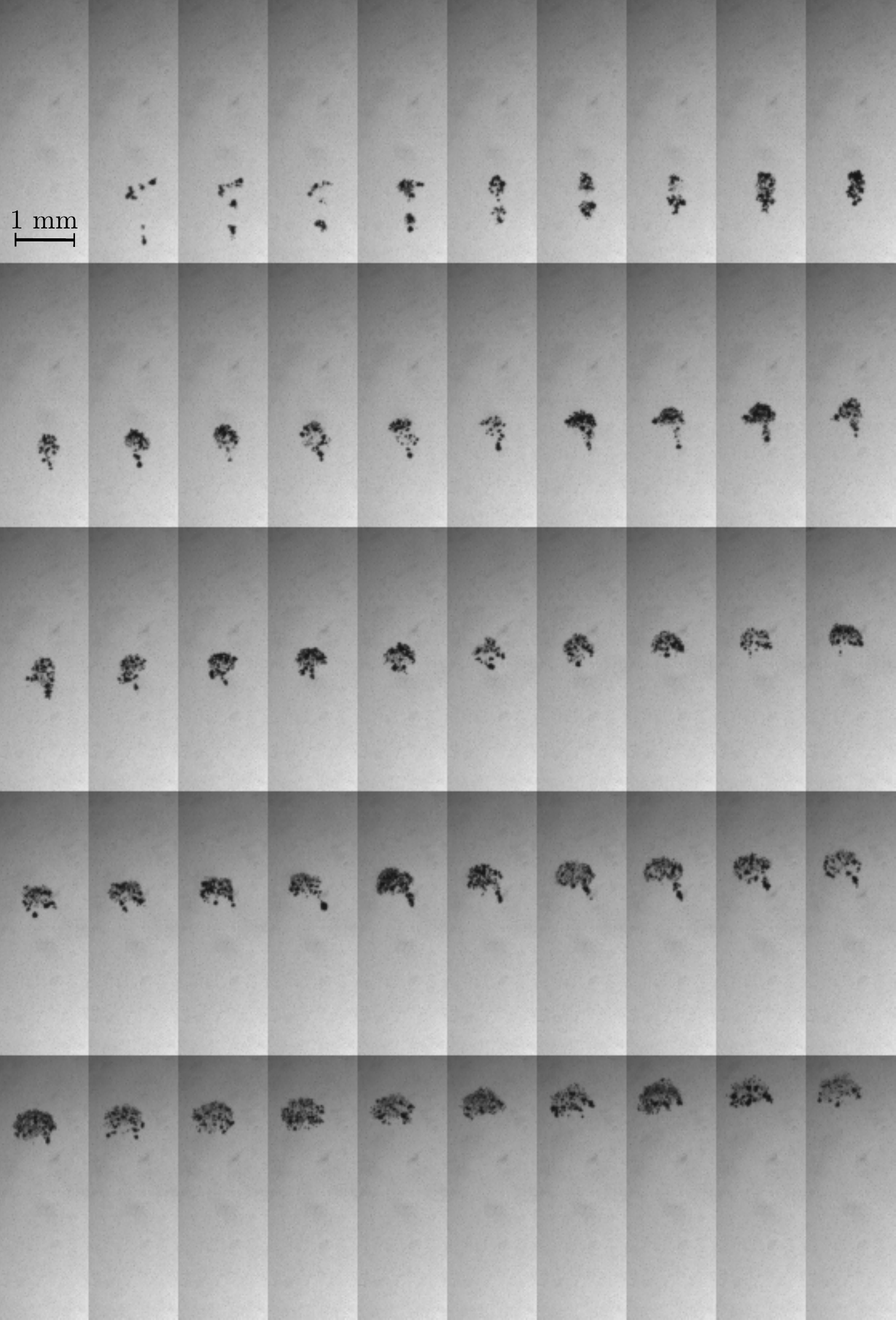} %reprint
    \caption{Time sequence of cavitation bubble cloud in a 250~\kilo\hertz~HIFU field. The first frame corresponds $t=0$. The interframe time is 40~\micro\second.}
    \label{fig:TimeSequenceCloud}
\end{figure}

We first consider the cavitation nucleation and the propagation of the resulting bubble cloud in the absence of a stone. Figure~\ref{fig:TimeSequenceCloud} shows the time history of a bubble cloud over 2~\milli\second~of ultrasound excitation. 

Multiple subclouds (at least five) first cavitate on different nucleation sites before merging to form a single, larger bubble cloud. This latter eventually travels upwards, in the direction of the ultrasonic wave. Upon merging, the cloud mostly adopts a spherical shape within a short duration ($\approx$ 10-15 acoustic periods) and then deforms into an umbrella-like shape (see a typical umbrella-shaped cloud on the penultimate frame on the second row in Fig.~\ref{fig:TimeSequenceCloud}). The umbrella shape consists in a spherical cap with a tail of bubbles. The umbrella shape lasts over 300 acoustic periods before the tail disappears. The remaining spherical cap continues to propagate while expanding its spanwise diameter. It eventually forms a sheet of dispersed bubbles (possibly an annular structure, but shadowgraphs do not allow for an exact description of this final shape). The rise speed of the bubble cloud is constant and found to be 1-1.5~\meter\per\second. Both the umbrella shape and the rising speed of the cloud are consistent with the rare previous observations of Ohl \emph{et al.}~\citet{ohl2015bubbles}. 

\begin{figure}[htbp]
    \centering
    \includegraphics[width=.5\columnwidth]{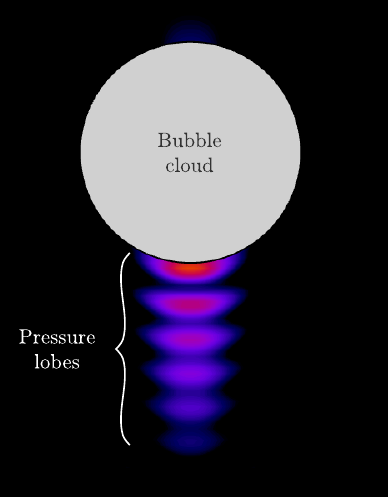} %reprint
    \caption{Acoustic simulation of the pressure field in the presence of a spherical bubble cloud, represented by a single bubble.}
    \label{fig:umbrella}
\end{figure}

The underlying physics behind the umbrella shape has not yet been elucidated and, unfortunately, our experimental data are not sufficient to address this morphology. It has been hypothesized that the tail below the spherical cap is a nucleation site that continuously produces bubbles, which eventually forms a tail by migrating towards the head of the cloud. However, the fact the tail suddenly disappears in a region where cavitation is likely to occur while the transducer is still running does not support this hypothesis. Based on acoustic simulations, we suggest an alternative candidate to explain the umbrella shape. In agreement with previous works~\cite{maeda2019bubble}, the scattering of the acoustic energy at the interface of a dense, spherical bubble cloud, here modeled as a single bubble, generates aligned pressure lobes as seen in Fig.~\ref{fig:umbrella}. The envelope of the bubble cloud and the pressure lobes are similar in shape with the umbrella. Assuming that the pressure inside the lobes exceeds the pressure threshold for cavitation nucleation (or rather, the tension), an umbrella-like shape cloud could be formed. During the upwards propagation of the cloud, the spherical head of the cloud becomes a sheet of dispersed bubbles that experiences a radial expansion, as seen and described in Fig.~\ref{fig:TimeSequenceCloud}. At this stage, one could suspect the scattering to occur at the interface of the individual bubbles constituting the cloud, and not at the cloud interface longer exist. 

\begin{figure}[htbp]
    \centering
    \includegraphics[width=\columnwidth]{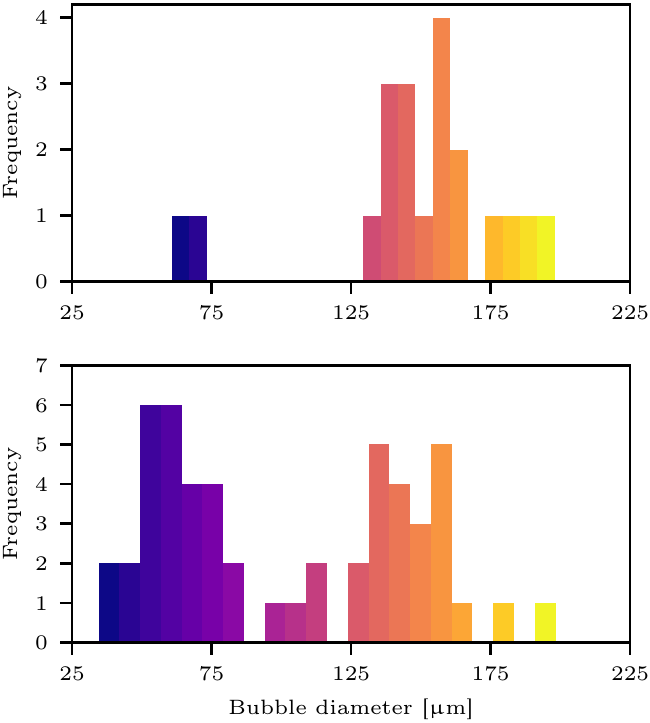} %reprint
    \caption{Bubble size distribution typically observed for a bubble cloud nucleating in the focal region. (Top) First frames where bubbles are observed. 75\% of the detected volumes have been identified as individual bubbles and used to plot the distribution (22 bubbles). (Bottom) Bubble size distribution $1.5$ acoustic periods later. 60\% of the detected volumes have been identified as individual bubbles and used to plot the distribution (53 bubbles).}
    \label{fig:distribution}
\end{figure}

Performing spatially resolved shadowgraphy by means of a long-distance microscope objective (Keyence, VH-Z50L) mounted on the high-speed camera enables to characterize the bubble size distribution and the bubble number density, $n$, defined as the ratio of number of bubble per unit volume. Figure~\ref{fig:distribution} shows typical distributions observed at two different times: the top graph corresponds to the first frame where bubbles are observed (\textit{i.e.}, early stage of the nucleation and bubble cloud formation) and the bottom graph to 1.5 acoustic periods later. Interestingly, the bubble size distribution initially has a negative skew with a peak around 150~\micro\meter~and later becomes bimodal with an additional peak at around 50~\micro\meter. This suggests that the first acoustic wave of the ultrasound burst generates vapor bubbles of relatively monodisperse size distribution, and that additional bubbles cavitate when the successive waves reflect at the surface of the initial bubbles. This would also explain the phase shift in the bubble oscillations between different bubbles within the cloud which can be observed in Fig.~\ref{fig:TimeSequenceCloud}. With regards to the bubble number density computed from the various imaged clouds, at the early stage before the bimodal distribution shapes, we found $n=32.6\pm7.3~\milli\meter^{-3}$.

\begin{figure}[htbp]
    \centering
    \includegraphics[width=\columnwidth]{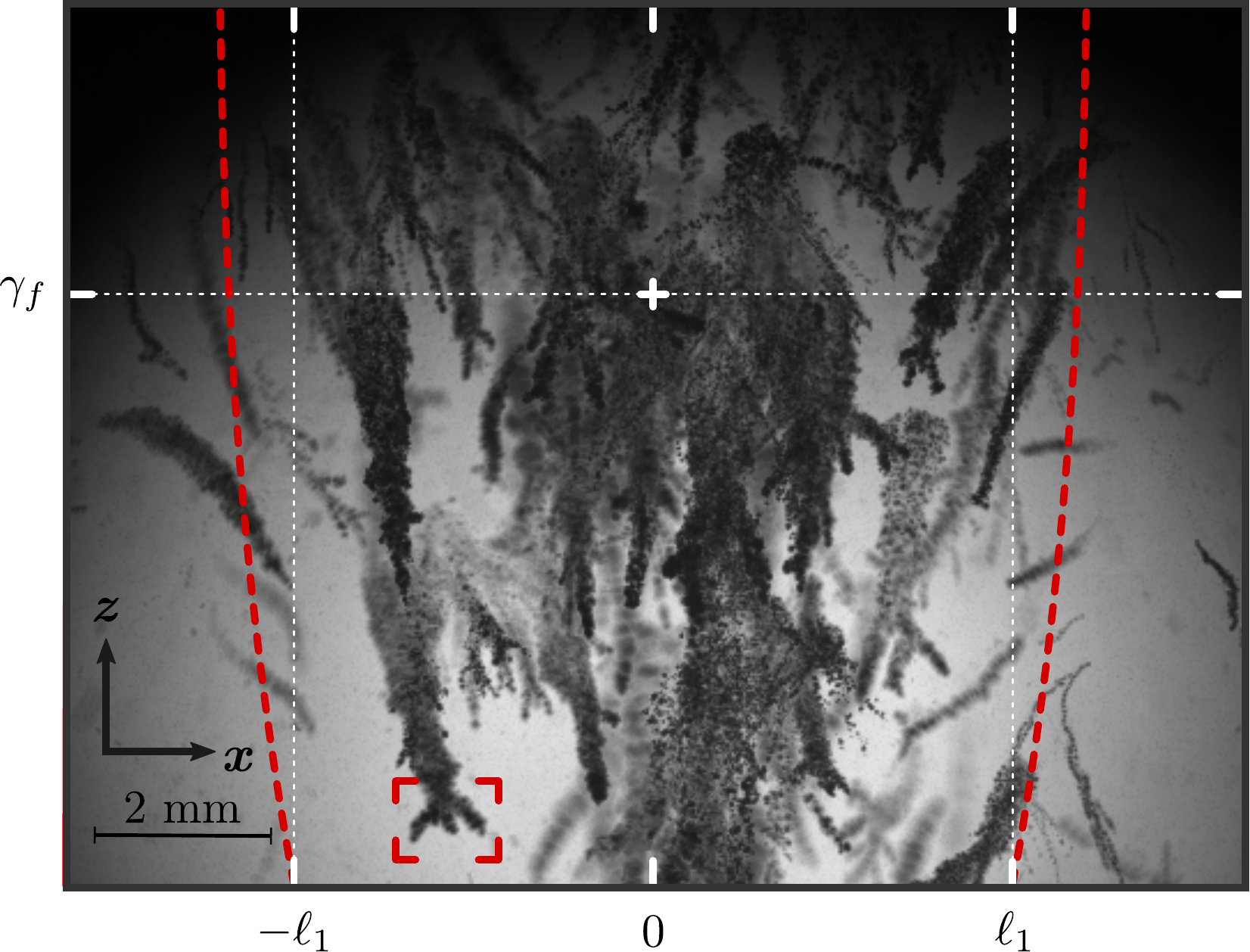} %reprint
    \caption{Minimum intensity projection from a single experiment showing hundreds of ultrasound-driven bubble clouds in a stone-free environment. The camera sampling frequency is 50~\kilo\hertz. The 250\kilo\hertz-transducer operated in burst mode at 10\% duty cycle (500 cycles, 2~\milli\second). The present projection has been compiled from the first 18~\milli\second. The dashed red lines contours the pressure region where $p^{*}>p_c^*$ as extracted from the numerical simulation.}
    \label{fig:CompiledClouds}
\end{figure}

Figure~\ref{fig:CompiledClouds} displays a MinIP-compiled image of the 18~\milli\second~of ultrasound excitation using the 250\kilo\hertz-transducer operating in burst mode at 10\% duty cycle (500 cycles, 2~\milli\second). Such reconstructed images illuminate the trajectory traced out by the single bubble clouds. As seen in the red frame, most of the trajectories starts with a dendritic structure drawn by the merging process of the subclouds. The mean rise speed of the hundreds of clouds captured in Fig.~\ref{fig:CompiledClouds} is 1.32~\meter\per\second. The highest density of clouds is reported near the acoustic focus $\gamma_f$. Interestingly, the envelope of the trajectories is a diverging beam which adapts to the shape of the ellipsoidal focal region, as seen in Fig.~\ref{fig:ExpVsNum}, and where $p^{*}>0.46 p^{*}_\mathrm{max}$ (see red dashed lines). This suggests $p_c^*\approx0.46 p^{*}_\mathrm{max}$ to be the approximate cavitation threshold for the present experiments. While we consider positive pressures throughout this study, it should be noted that it is the tensile component of the ultrasound wave that generates cavitation.
%
% *************  Bubble-cloud vs stone ************

\subsection{Bubble cloud near a model stone}

\begin{figure}[htbp]
    \centering
    \includegraphics[width=\columnwidth]{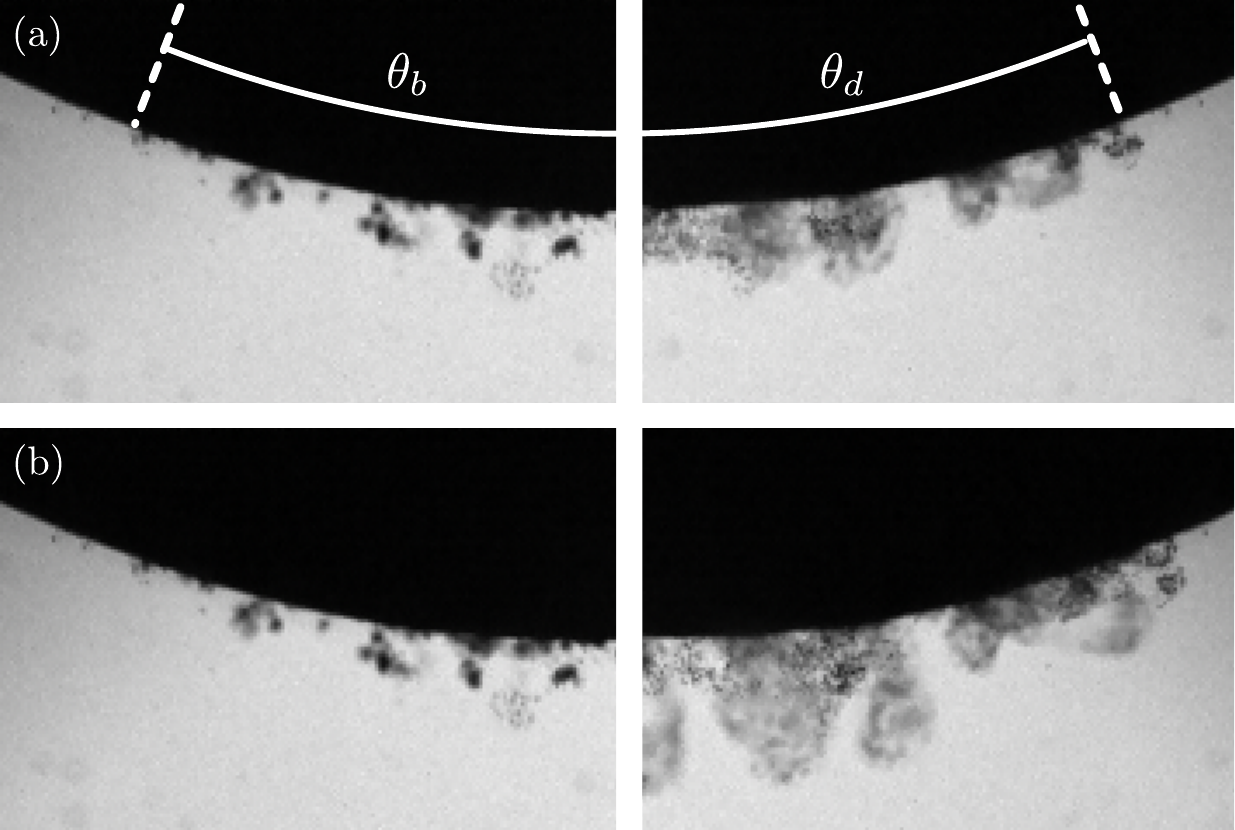} %reprint
    \caption{[(a-b), left panels] Interaction of bubbles clouds with an hemispherical stone of 20 \milli\meter~in diameter 20~\micro\second~after the ultrasound burst starts. [(a-b), right panels] Falling fragments of stone captured 4.8~\milli\second~and 18.8~\milli\second~after the end of the ultrasonic excitation.}
    \label{fig:DamageVsCloud}
\end{figure}

Figure~\ref{fig:DamageVsCloud}(a-b) are mirrored MinIp-compiled images where the left-hand sides show the initial bubble cloud distribution near the stone surface at $t=0~\milli\second$ (burst on), and the right-hand sides display the resulting falling stone fragments, after the ultrasound burst is off, for time (a) $t=5~\milli\second$ and (b) $t=15~\milli\second$. The stone diameter is 20~\milli\meter~with the south-pole located in $(r,\gamma_f)=(0,1)$. The normal projection of the bubble cloud layer on the stone surface shapes a spherical cap whose edge is located at $\theta_b$ (angle from the vertical axis), so that the cap area reads $\mathcal{A}_b=2\pi R^2(1-\cos\theta_b)$. Similarly, the edge of the normal projection of the falling fragments on the stone surface is located at $\theta_d$ and its area equals $\mathcal{A}_d=2\pi R^2(1-\cos\theta_d)$. Independently of the stone size and $z$-distance to the transducer, experiments report $\theta_b\approx\theta_d\Leftrightarrow\mathcal{A}_b\approx\mathcal{A}_d$. This suggests that the bubble hammers the stone surface by collapsing and jetting against it.

\begin{figure}[htbp]
    \centering
    \includegraphics[width=\columnwidth]{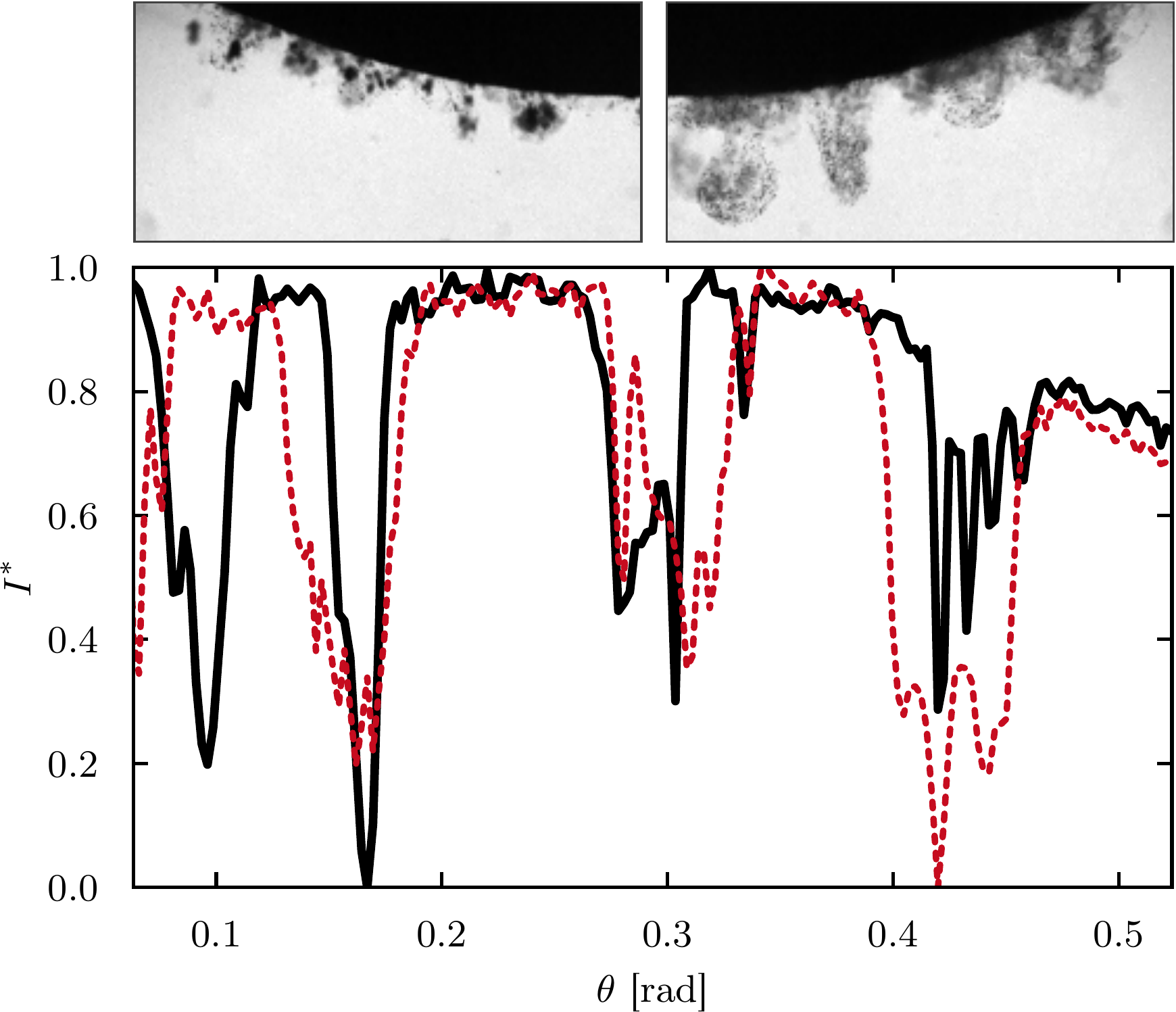} %reprint
    \caption{Measurement of the normalized pixel intensity $I^{*}$ over the polar coordinate $\theta$ given in radian. The solid dark line and the red dashed lines have been extracted from the top left and right images, respectively. The left image is 60~\micro\second~after the ultrasound burst starts, and the right image is 23.6~\milli\second~later. Intensity fluctuations result from bubbles or stone fragments along the optical path.}
    \label{fig:PlotDamageVsCloud}
\end{figure}

 Furthermore, when plotting the pixel intensity fluctuations along the stone surface (see Fig.~\ref{fig:PlotDamageVsCloud}) in both the bubbly and dusty regions, it appears that the polar location of the bubble clouds matches very well with the location of the clusters of falling fragments. In the discussion whether the bubbles contribute, or not, to the stone comminution (in addition to the acoustic energy), this result supports the direct role of cavitation bubbles in the stone fragmentation, at least at the surface erosion level. Speculating on an exclusive contribution of the acoustic wave energy in the stone destruction, one would expect the fragments to form a homogeneous layer rather than the plume structures as observed on the top right image in Fig.~\ref{fig:PlotDamageVsCloud} and the right image in panels (a) and (b) of Fig.~\ref{fig:DamageVsCloud}. Note that the stone fragments are ejected normal to the stone surface (see the preferential direction of the plume structures), which is consistent with the bubble activity described earlier. One could suspect the bubble hammering to create local pits from where fragments are eventually ejected. Furthermore, similarly to cavitation nucleation, the damage area seems to follow more of a thresholding behaviour with pressure rather than direct proportionality that one could expect with direct pressure-induced damage.
 
\begin{figure}[htbp]
    \centering
    \includegraphics[width=\columnwidth]{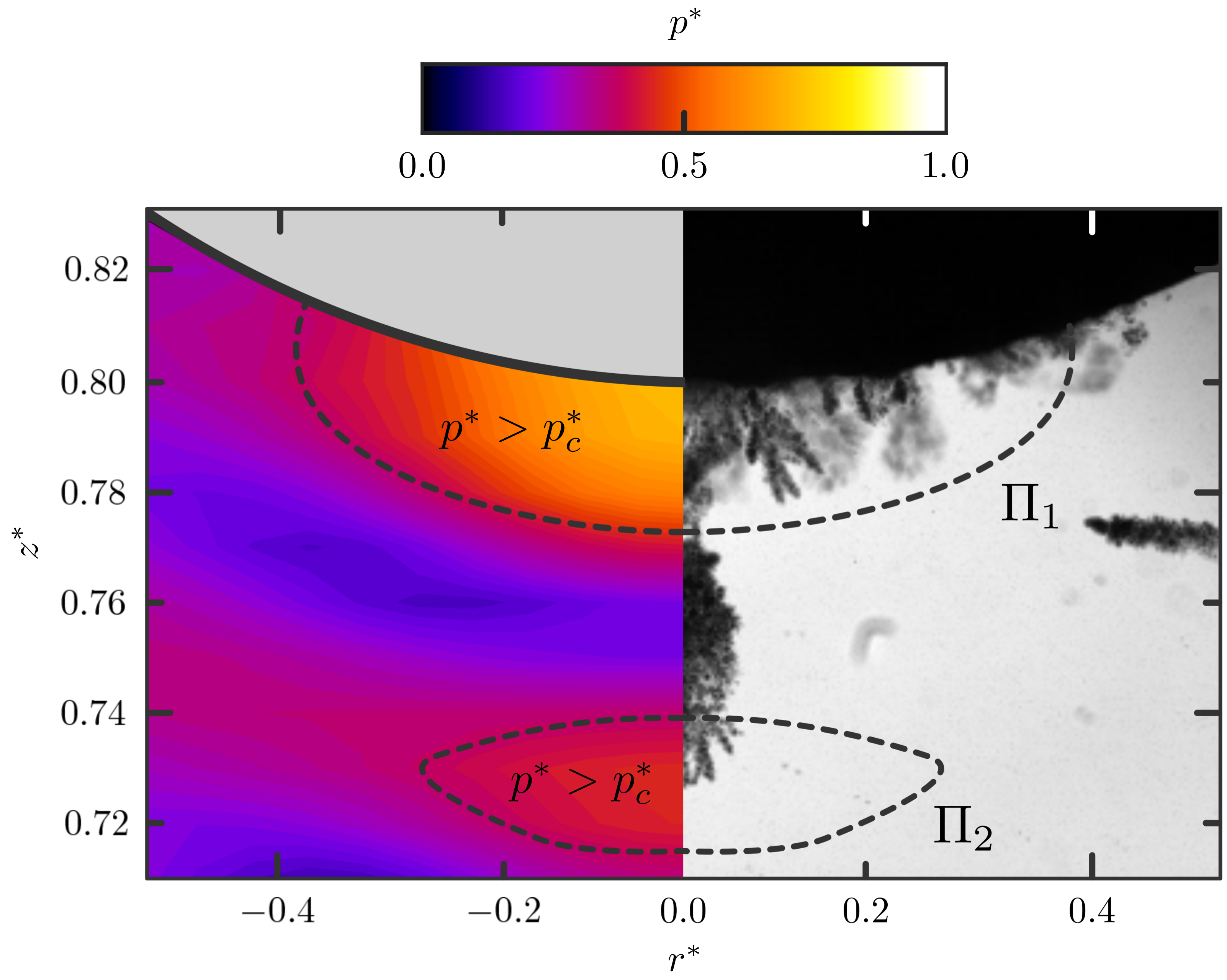} %reprint
    \caption{Comparison of the pressure field computed from the acoustic simulations against the MinIp-compiled shadowgraph of the corresponding experiment run over one burst. The stone is a 20-\milli\meter~diameter hemisphere. Regions $\Pi_1$ and $\Pi_2$ locate the high-pressure regions where the pressure cavitation threshold is reached, $p^{*}>p_c^*$.}
    \label{fig:BegoStone}
\end{figure}

However, comparing the bubble cloud density captured in Fig.~\ref{fig:CompiledClouds} and Fig.~\ref{fig:DamageVsCloud}, much fewer clouds are generated in the presence of a stone, and the few clouds generated do not shape the ellipsoidal profile of the focal region as previously observed (where $p^{*}>p_c^*$). To investigate such discrepancies, acoustic simulations are performed and results compared against the experimental observations. Figure~\ref{fig:BegoStone} shows the interaction of the HIFU field with a half-sphere stone. The computed acoustic pressure field shows that the stone scatters the incoming wave which prevents the formation of the ellipsoidal focal region, but instead creates a pattern consisting in pressure lobes. The pressure pattern thus results in successive high pressure layers whose magnitude decreases with $z$. The pressure regions where $p^{*}>p_c^*$ are contoured using dark dashed lines. A direct comparison with the experimental shadowgraphs, compiled using the MinIP-algorithm, shows that the cloud nucleation sites are located within these regions. This latter is consistent with observations provided on Fig.~\ref{fig:CompiledClouds}. Figure~\ref{fig:BegoStone} also shows that only clouds that nucleate within the $\Pi_1$ high-pressure region, attached to the stone surface, eventually reach the stone surface and are able to generate damage. Clouds that cavitate in the isolated high-pressure regions further away from the stone ($\Pi_2$-region) are trapped in the acoustic field in-between the high-pressure $\Pi_1$ and $\Pi_2$-regions. Bubbles in between the $\Pi_1$ and the $\Pi_2$ region result from translation, not nucleation. Note that, in the following, we denote $\theta_p$ the edge of the spherical cap formed by the normal projection of the $\Pi_1$-region on the stone surface and $\mathcal{A}_p$ its area, which equals to $2\pi R^2(1-\cos\theta_p)$.

\begin{figure}[htbp]
    \centering
    \includegraphics[width=\columnwidth]{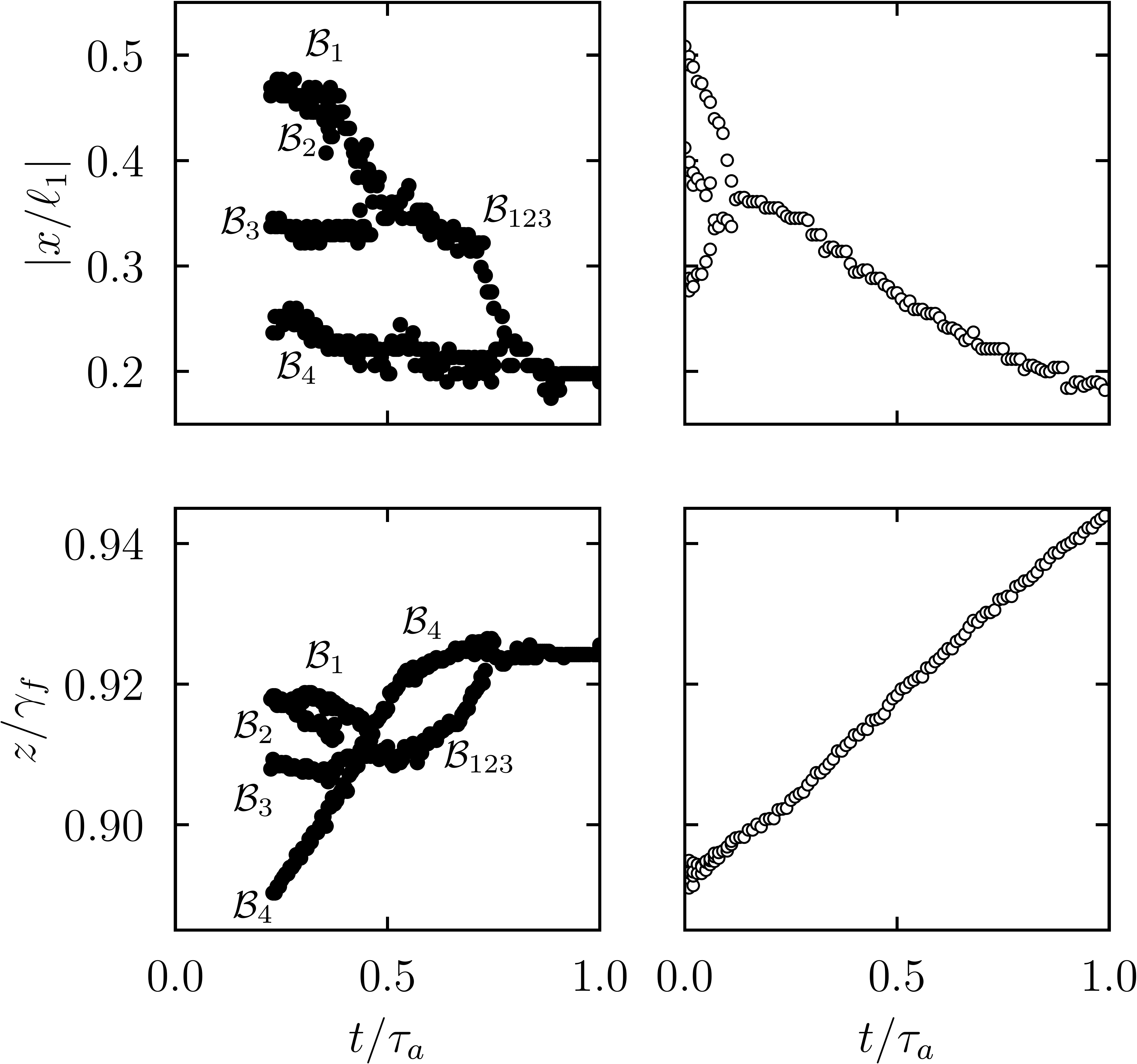} %reprint
    \caption{Trajectory of bubble clouds along the $x$ and $z$-direction as time proceeds. The physical time is normalized by the acoustic period $\tau_a$. The two left graphs correspond to experiments with a stone, and the two right plots correspond to bubble clouds propagating in a stone-free environment. The $\mathcal{B}_i$ with $i=1,2,3,4$ refer to the initial subclouds. The cloud denoted $\mathcal{B}_{123}$ is formed from the merging of $\mathcal{B}_1$, $\mathcal{B}_2$ and $\mathcal{B}_3$. }
    \label{fig:CloudTraj}
\end{figure}

Figure~\ref{fig:CloudTraj} (left-hand side, top and bottom) shows the trajectory of bubble clouds that initially nucleate in the $\Pi_2$-region. Upon turning on the ultrasonic excitation, four subclouds are generated (we refer to these subclouds as $\mathcal{B}_i$ with $i=1,2,3,4$). Subclouds $\mathcal{B}_1$ and $\mathcal{B}_2$ first merge to form $\mathcal{B}_{12}$, which eventually merges with $\mathcal{B}_3$ around time $t/\tau_a=0.45$, where $\tau_a$ is the acoustic period, and forms $\mathcal{B}_{123}$. This merged cloud propagates towards $\mathcal{B}_4$ and meets it around $t/\tau_a=0.75$. In the end, for time $0.75<t/\tau_a<1.0$, a single cloud resulting from the four subclouds remains. According to both the $x$ and $z$-location, this latter cloud stagnates in the acoustic field, that is, its translational motion is zero. This behavior is characteristic of the presence of a stone in the acoustic path. The incident beam is reflected at the stone surface. The reflected wave superposes with the incident wave, which creates a standing wave. The bubble cloud generated within the standing wave thus propagates towards nodes where they are trapped, i.e., they levitate. We observe that when clouds cavitate sufficiently far from the stone surface, they first stagnate in the acoustic field before travelling in the opposite direction of the stone. This corresponds to the very short time when no more incident waves propagate but the reflected wave is still established.
Figure~\ref{fig:CloudTraj} (right-hand side, top and bottom) shows the trajectory of a bubble cloud in the absence of a stone. Upon total merging of the subclouds, and under the action of the primary Bjerknes forces (acoustic radiation forces, i.e., translational force resulting from the non-zero time-averaged acoustic pressure gradient due to the volumetric bubble oscillations~\cite{leighton1990primary}), the merged bubble cloud follows a linear trajectory with constant speed (vertical speed, $u_z\approx1.57~\meter\per\second$). 
Note that we assume that gravity plays no role, as well as the acoustic streaming~\cite{lu2013dynamics,willard1953ultrasonically}. By contrast, the rising speed of $\mathcal{B}_4$ for $t/\tau_a\in[0,0.5]$ is $u_z\approx2.70~\meter\per\second$, which is 2.05 more than the mean speed computed from Fig.~\ref{fig:CompiledClouds}. Rise speeds of the $\mathcal{B}_{i}$ clouds are given in Table~\ref{tab:table1}.
\begin{table}[htbp]
\caption{\label{tab:table1}%
Velocity of the bubble clouds according to the bottom left plot in Fig.~\ref{fig:CloudTraj}. $\mathcal{B}_{1234}$ stands for the ultimate bubble cloud after total merging of all subclouds. The dimensionless speed $u_z^{*}$ is defined as $u_z$ normalized to the mean rise speed computed from Fig.~\ref{fig:CompiledClouds}.
}
\begin{tabular}{lccccr}
 & $\mathcal{B}_4$ & $\mathcal{B}_4$ & $\mathcal{B}_4$ & $\mathcal{B}_4$ & $\mathcal{B}_{1234}$\\
\hline
$u_z$ (\meter\per\second) & 0.92 & 1.02 & 0.39 & 2.07 & 0.019\\
$u_z^{*}$ ($-$) & 0.70 & 0.78 & 0.30 & 2.05 & 0.015\\
\end{tabular}
\end{table}

\begin{figure}
    \centering
    \includegraphics[width=\columnwidth]{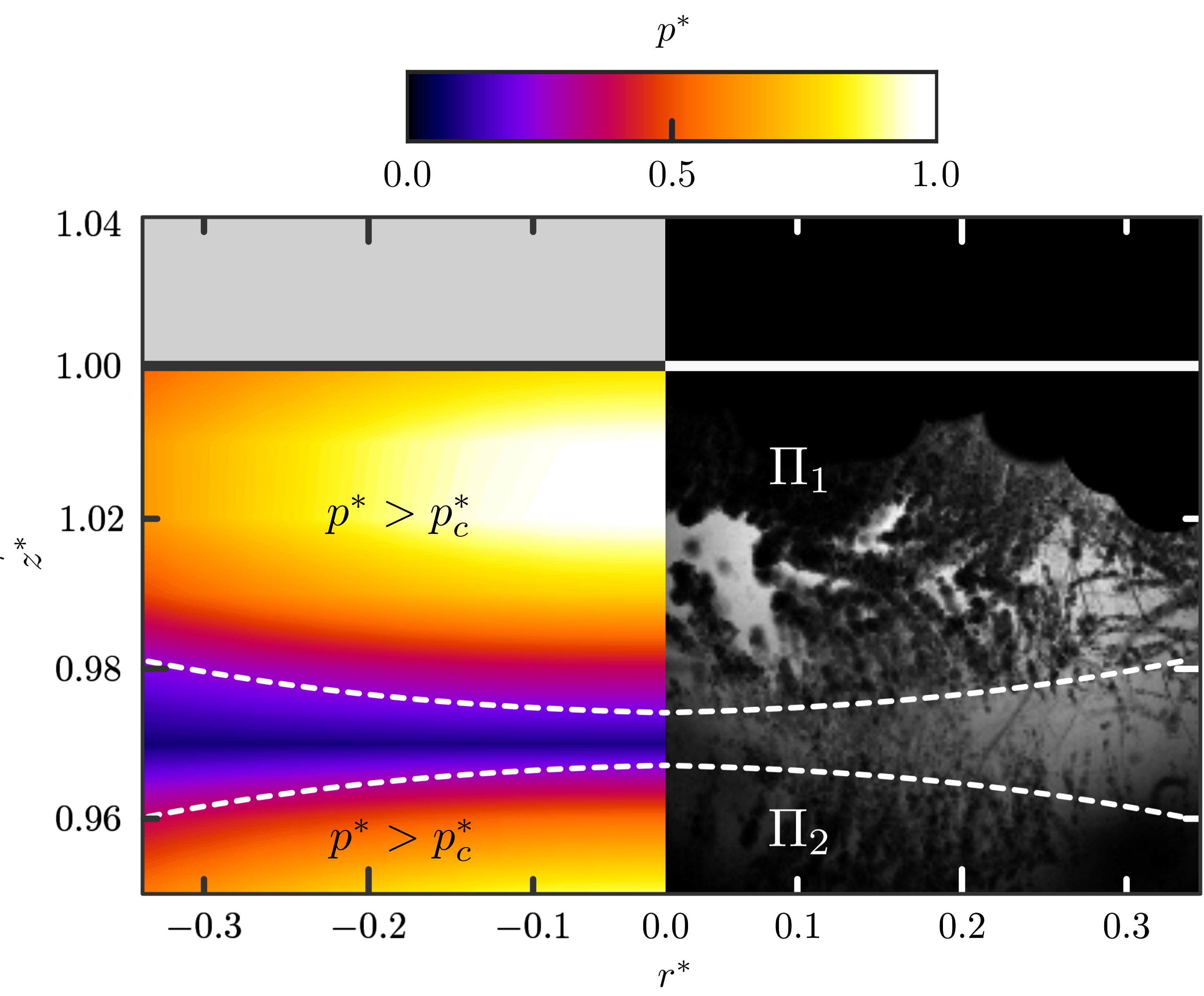} %reprint
    \caption{Comparison of the pressure field computed from the acoustic simulations against the MinIp-compiled shadowgraph of the corresponding experiment run over one ultrasound burst. The stone has a flat surface of 35-\milli\meter~side length. Again, regions $\Pi_1$ and $\Pi_2$ locate the high-pressure regions where $p^{*}>p_c^*$.}
    \label{fig:BlueStone}
\end{figure}

To assess the phenomenology derived from the comparison given with Fig.~\ref{fig:BegoStone}, we performed a similar simulation using a stone with a flat surface, as well as the corresponding experiment. Figure~\ref{fig:BlueStone} shows the comparison of the results. One can observe that only a few bubbles occasionally translate in-between the $\Pi_1$ and $\Pi_2$-regions, while hundreds of clouds are captured within these two domains. This indicates that no bubbles nucleate in this low-pressure region nor propagate through, which tends to confirm the phenomenology suggested earlier.

\subsection{Prediction of the optimum source-to-object distance} 
In the context of lithotripsy, maximizing the surface damage on the stone surface is of primary interest. Based on the previous observations that the damage happens in the area where cavitation nucleates near the stone, maximizing the damage area would consist in maximizing the spanwise diameter of the $\Pi_1$-region, which corresponds to the stagnation pressure region. This is achievable by optimizing the source-to-object distance. Based on the theoretical prediction of the on-axis pressure generated by the transducer, given by the O'Neil integral solution, and remembering that the pressure induced by the reflection at a boundary reads $p_r=2p_iI_2/(I_1+I_2)$ \cite{thompson1972compressible}, where $p_i$ is the incident pressure and $I_1$ and $I_2$ are the acoustic impedance of water and the stone, respectively, one can predict the on-axis pressure of the reflected wave. 
\begin{figure}
    \centering
    \includegraphics[width=\columnwidth]{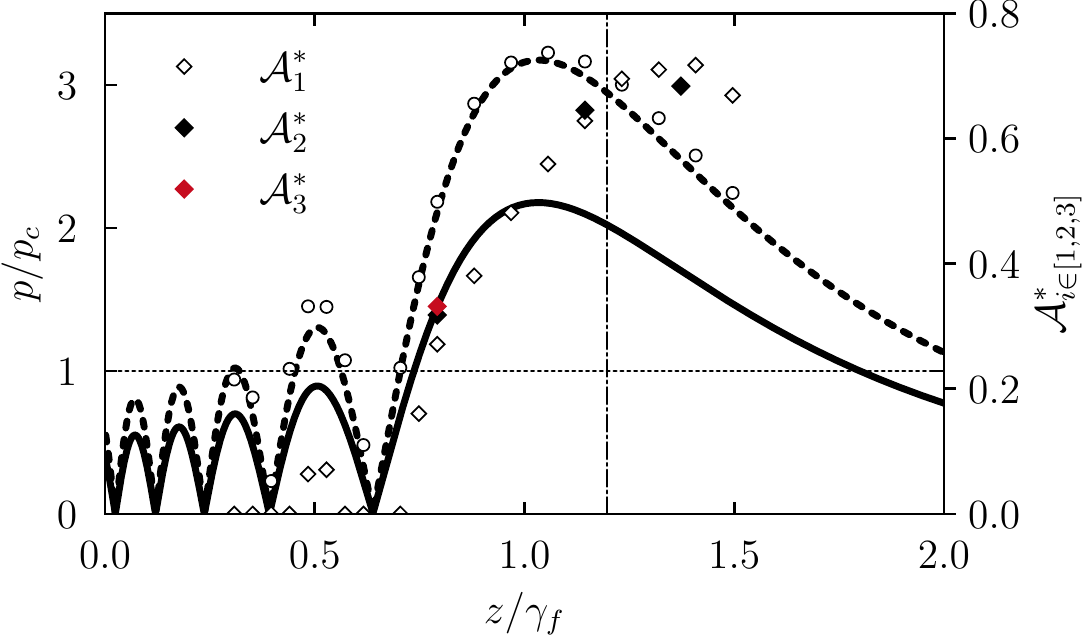} %reprint
    \caption{Axial pressure normalized to the cavitation pressure threshold, and surface damage coefficients over the $z$-direction. Surface damage coefficients are defined with Table~\ref{tab:table2}. The thick solid line corresponds to the O'Neil integral solutions, and the thick dashed line is the theoretical reflected pressure at stone surface. The circle markers shows the simulated reflected pressure. The dotted-dashed line is the geometrical focal distance. The dotted line plots the cavitation pressure threshold.}
    \label{fig:prediction}
\end{figure}
Figure~\ref{fig:prediction} plots the O'Neil integral solution (thick solid line) compared with the theoretical and the numerical $p_r$ solutions (thick dashed line and circle markers, respectively), and displays a good agreement between the theoretical prediction and the numerics. In addition, it reveals that even the precursor pressure lobes in the incident pressure profile (e.g., at $z/\gamma_{f}=0.5$) are likely to generate, upon reflection at the stone surface, high-pressure regions where cavitation may occur.

\begin{figure}[htbp]
    \centering
    \includegraphics[width=\columnwidth]{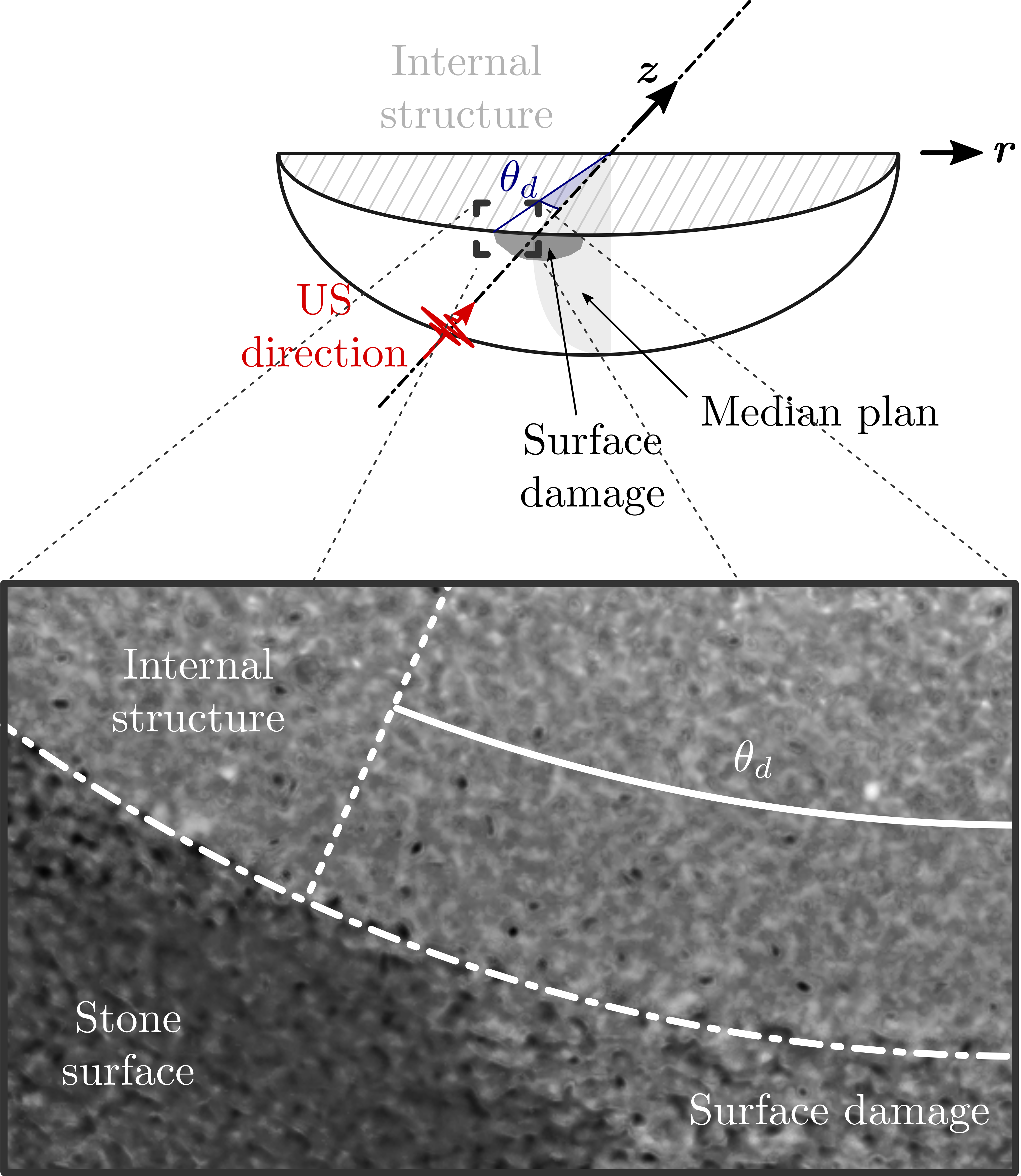} %reprint
    \caption{Tomography of the surface damages at the proximal pole of a 20-\milli\meter~diameter hemispherical model stone after being exposed to focused ultrasound. The 250\kilo\hertz-transducer was operated in burst mode at 10\% duty cycle (500 cycles, 2~\milli\second) during 2~\second. The tomography has been performed at the beamline ID19 of the European Synchrotron Research Facility. Additional details on the optical alignment of the tomograph are given in Appendix~\ref{app:a}.}
    \label{fig:TomoBego}
\end{figure}

In the following, we compare three surface damage coefficients,  $\mathcal{A}_{i\in[1,2,3]}^{*}$, representing the relative surface areas for i)~pressure beyond the threshold pressure computed numerically, ii)~cavitation bubble nucleation visible from shadowgraphs and iii)~surface damage determined from X-ray tomography. The tomograph is obtained using synchrotron X-rays and an example plane is shown in Fig.~\ref{fig:TomoBego}, while details on the methodology are provided in Appendix~\ref{app:a}. The exact definitions for each coefficient are provided in Table~\ref{tab:table2} where $\mathcal{A}_0$ is the initial area of the exposed surface of the stone ($=\pi R^2/2$). 
\begin{table}[htbp]
\caption{\label{tab:table2}%
Definitions of the surface damage coefficients depending on the method: (i) acoustic numerical simulation, (ii) experimental shadowgraph, and (iii) the experimental tomograph.
}
\begin{tabular}{lccr}
 Method & $\mathcal{A}_{i\in[1,2,3]}^{*}$ & Definition\\
\hline
Numerics & $\mathcal{A}_1^{*}$ & $\mathcal{A}_{p}/\mathcal{A}_{0}=4(1-\cos\theta_p)$\\
Shadowgraphs & $\mathcal{A}_2^{*}$ & $\mathcal{A}_{b}/\mathcal{A}_{0}=4(1-\cos\theta_b)$\\
Tomographs & $\mathcal{A}_3^{*}$ & $\mathcal{A}_{d}/\mathcal{A}_{0}=4(1-\cos\theta_d)$\\
\end{tabular}
\end{table}
Figure~\ref{fig:prediction} plots the $\mathcal{A}_{i\in[1,2,3]}^{*}$ in direct comparison with the theoretically and numerically computed pressure fields. Despite a single tomography-based measurement of the surface damage, the $\mathcal{A}_1^{*}$, $\mathcal{A}_2^{*}$ and $\mathcal{A}_3^{*}$ coefficients are found to be in very good agreement. This result indirectly evidences the contribution of cavitation bubbles in the stone surface erosion. In addition, the variation of $\mathcal{A}_1^{*}$ and $\mathcal{A}_2^{*}$ along the transducer axis shows that the optimum source-to-object distance, $z_{+}$, to maximize the stone surface damage area is located in $\gamma_f+\ell_2/2$. This corresponds to the center of the ellipsoidal focal region in the $z$-direction. While intuitive, this optimum distance is only true if $p_r(z=\gamma_f+\ell_2/2)>p_c$. To be more general, the optimum distance ranges as $\gamma_f<z_{+}<\gamma_f+\ell_2/2$ depending on the magnitude of the incident pressure field. More specifically, 

\begin{equation}
    \begin{cases}
    z_{+} = \gamma_f ~~~~~~\qquad \mathrm{if} \quad p_i(\gamma_f)= p_c,\\
    z_{+} = \gamma_f+\ell_2/2 \quad \mathrm{if} \quad p_i(\gamma_f+\ell_2/2)\geq p_c.
    \end{cases}
    \label{eq:limz}
\end{equation}

It should be noted that this corresponds to the optimum for the damage area. What is not investigated here is the damage depth and the contribution to the final fracture of the entire stone, which should be addressed in future studies.

\subsection{Discussion on implications}

The presented results shed light on the location and extent of cavitation activity generated by a laboratory-scale HIFU transducer both in the bulk as well as in the presence of a model stone. 
It should be noted that nucleation characteristics in both scenarios are highly sensitive to parameters that we do not systematically control in this study, such as fluid and surface properties.
Here we use tap water of which the relatively high level of air saturation and impurity content facilitate cavitation nucleation.
On the other hand, the nucleation of cavities on the surface of the model stone depends on the surface microgeometry and the state of wetting.
Nonetheless, what we learn from the presented results on the fundamental physics of acoustic cavitation are i)~how nucleation locations, trajectories and cloud characteristics in focused acoustic fields are linked to the acoustic pressure fields, as demonstrated both within the bulk and in the presence of two differently shaped acoustically reflective objects, and ii)~insights to explain the evolution of the bubble cloud shapes, such as the umbrella and cloud shapes as well as changes in bubble size distribution within the cloud.

What we have learned with respect to the contribution of cavitation bubble clouds on the model stone damage, on the other hand, is that the area of the surface with observable erosion, which is measured both through visual observations of fragments and X-ray tomography quantifying the material removal, closely correlates with the visualised cavitation activity, even when the cavitation bubbles are not directly attached to the stone surface (see Figure~\ref{fig:PlotDamageVsCloud}). The mechanisms in play may, however, differ from the stronger acoustic fields associated to the full-scale medical lithotripters where a full layer of vapor bubbles may cover the entire stone surface.
This layer is believed to be responsible for the main stone damage but also shields off subsequent acoustic cycles.
However, these findings can be valuable in the effort to understand scaling effects when considering cavitation contributions in acoustic fields produced by full-scale transducers relevant for medical lithotripsy.
Furthermore, the implications of our results for the cavitation bubble clouds in the bulk may be useful in histotripsy (where HIFU is used to make surgical cuts in biological tissue) or other applications using focused ultrasound to produce cavitation bubble clouds in different media where the control of their characteristics is desired.

% **********************************************************
% *************       Conclusion        ********************
% **********************************************************

\section{Conclusion}
We investigated the formation and role of cavitation bubble clouds in the surface damage of a model stone in a lab-scale high-intensity focused ultrasound field. We first described the bubble cloud properties and trajectories in a stone-free bulk liquid. In agreement with previously published work, we observed umbrella-shaped clouds propagating upwards at a constant mean speed of 1.5~\meter\per\second. We complemented these results with original findings on the size distribution and bubble number density. The size distribution initially exhibits a single peak around 150~\micro\meter~for an approximate mean number density of $n=33~\milli\meter^{-3}$. The bubble size distribution change for a bimodal profile with peaks at 50 and 150~\micro\meter, suggesting that secondary cavitation bubbles appeared upon reflection of the acoustic wave at the surface of the initial bubbles. Experimental observations also show that the cloud trajectories adopt the acoustic beam shape, thus diverging in the south pole of the acoustic focal region. Combining experimental and numerical results, we speculated the umbrella shape to be related to the scattering of the acoustic energy at the bubble cloud interface. However, significant research efforts are still required to confirm this hypothesis. 

Positioning a model stone in the focal region of the transducer, we correlated the bubble activity of the damages reported at the stone surface. We demonstrated that the location of the fragments captured with shadowgraphs, and the surface erosion observed from tomographs, matches with the bubble-stone contact sites. Based on acoustic simulations and direct comparison against experimental observations, we identified the nucleation sites which contribute to the stone erosion. We eventually defined three surface damage coefficients to evaluate the optimum source-to-object distance, $z_+$, which, depending on the pressure magnitude, has been found to range from $\gamma_f$ to $\gamma_f+\ell_2/2$.

Future works would now consists to extend such an analysis to transducers and lithotripters with acoustic amplitudes and energies that are directly relevant to the respective applications, and to consider model stones with relevance in terms of geometry, structures and composition. The relation between surface damages and damage depth, over cycles, is of major interest and still requires research efforts. Also, the transition from surface damages to cracks and break-up should be addressed to assess the efficiency of the different ultrasound-based stone-breaking techniques.

% **********************************************************
% *************       Acknowledgements        **************
% **********************************************************

%% before appendix (optional) and bibliography:
\noindent\textbf{Acknowledgments}
The authors acknowledge the financial support from the ETH Zurich Postdoctoral Fellowship program. We also acknowledge the beamline ID19 of the European Synchrotron Radiation Facility where the tomograph has been performed as part of the allocated proposal beamtime ME-1599 on beamline ID19. Authors acknowledge Richard Guinchan and Claire Bourquard for their contribution to the characterization of acoustic pressure field.

% **********************************************************
% ******************       Appendix        *****************
% **********************************************************

% ---------------------------------------------------------------------
% Appendix  (optional)

\appendix
\section{Acoustic focus: theoretical background}
\label{app:a}

Focused ultrasound can be generated directly from a spherical transducer or by means of an acoustic lens which focuses the beam originating from a plane vibrating surface. In the latter case, two focal distances must be defined: (i) the geometrical focal distance $f$ which corresponds to the radius of curvature of the lens, and (ii) the acoustic focus $\gamma_f$ where the maximal pressure in the ultrasound beam is located. 

Let us consider the transducer and the acoustic lens to be aligned along the axial $z$-direction. The transducer is located in $z=0$ and the lens in $z=z_0$. The radius of the transducer is $a$. The position vectors $\mathbf{r}$ and $\mathbf{r^{\prime}}$ locate the field and source points, respectively. Accordingly, the radial coordinate of the field and source points are $r$ and $r^\prime$. The azimuthal angle is denoted $\varphi$ in the observation plane, and $\varphi^{\prime}$ in the plane of the lens.

Assuming a known pressure distribution $p(r,z_0)$, the pressure field $p(r,z)$ can be computed from the Rayleigh-Sommerfeld diffraction integral
\begin{equation}
    p(r,z)=\frac{-ik\tilde{z}}{2\pi}
    \iint_\Gamma
    \frac{p(r^\prime,z_0)\mathrm{e}^{ik}(|\mathbf{\tilde{r}}|^2+\tilde{z}^2)^{1/2}}{|\mathbf{\tilde{r}}|^2+\tilde{z}^2}r^\prime dr^\prime d\varphi^\prime,
    \label{eq:1}
\end{equation}
where $k$ is the wavenumber defined as the ratio of the angular frequency $\omega$ and the sound speed in the medium $c_0$, $\mathbf{\tilde{r}}=\mathbf{r}-\mathbf{r^{\prime}}$ and $\tilde{z}=z-z_0$. The time dependency given by the propagation term $\mathrm{e}^{-i(\omega t-k z)}$ is omitted. Modifying the Rayleigh integral with the Fresnel approximation, and using the identity $\mathbf{\tilde{r}}=r^2+{r^{\prime}}^{2}-2r r^\prime\cos(\varphi-\varphi^{\prime})$, the dimensionless radiated pressure field can be expressed by
\begin{equation}
    \bar{p}(\xi,\gamma)=\frac{2\pi}{i\gamma}
    \int_{0}^{\infty}
    \exp\left\{i\pi\frac{\xi^2+{\xi^{\prime}}^2}{\gamma}\right\}J_0\left[\frac{2\pi\xi\xi^\prime}{\gamma}\right]
    \bar{q}(\xi^\prime)\xi^\prime d\xi^\prime,
    \label{eq:2}
\end{equation}
where $\xi=r/a$, $\xi^{\prime}=r^{\prime}/a$, $\gamma=\tilde{z}/a$ and $\bar{q}(\xi^\prime)$ is the normalized source function. Note that the integral over $\varphi^\prime$ in Eq.~(\ref{eq:1}) has been replaced by the 0th-order Bessel function $J_0\left[{2\pi\xi\xi^\prime}/{\gamma}\right]$. Attenuation effects are not considered here. Considering a uniform ultrasonic excitation focused spherically, the normalized source function can be written as  
\begin{equation}
    \bar{q}(\xi^{\prime})=f(\xi^{\prime})\mathrm{e}^{-i\pi{\xi^{\prime}}^2/\beta}
    \quad \mathrm{with} \quad
    f(\xi^{\prime})=
    \begin{cases}
    1,&0\leq\xi^{\prime}<1,\\
    0,&\xi^{\prime}>1,
    \end{cases}
    \label{eq:3}
\end{equation}
where $\beta=f\lambda/a$ with $\lambda$ the source wavelength. The on-axis pressure amplitude, near the geometrical focus, eventually reads
\begin{equation}
    |\bar{p}(\xi^{\prime})|=\frac{2\beta}{\beta-\gamma}\sin\frac{\pi(\beta-\gamma)}{2\beta\gamma}.
    \label{eq:4}
\end{equation}
Differentiating Eq.~(\ref{eq:4}) with respect to $\gamma$ and setting the result equal to zero gives $\tan(\pi x)=\pi x(1+2\beta x)$, where $x=(\beta-\gamma)/(2\beta\gamma_f)$ and $\gamma_f$ is the axial coordinate of the physical focal distance, i.e., the axial coordinate of the maximum pressure. Approximating $x\approx\beta(1+2\beta)^{-1}$, one finds the acoustic focus to be at $\gamma_f \approx \beta[1+{2\beta^2}/({1+2\beta})]^{-1}$. Interested readers can find more details on these derivations in refs.~\cite{cavanagh1981lens,goldstein2006steady,huang2009simple}.

\section{Synchrotron X-ray tomography}
\label{app:b}

The \emph{operando} synchrotron X-ray microtomography was performed at the 145-m long imaging beamline ID19 of the European Synchrotron Radiation Facility (ESRF), Grenoble \cite{weitkamp2010status}. A pink beam (filtered with 2.2-mm Diamond, 2.8-mm Aluminium, 1.4-mm Copper, 0.06-mm Tungsten and successive thin carbon and beryllium windows,
mounted along the vacuum flight tube) with 68~keV peak energy, generated by a wiggler insertion device ($\lambda=15$~cm, gap 60~mm) was used to illuminate the sample. The beam was focused onto the sample using two pairs of in-vacuum slits positioned at 30~m (principal slits for heat moderation), and at 130~m (secondary slits for beam collimation). 
The setting provides sufficient transmission-to-flux compromise to reach relatively low exposure times. The indirect X-ray detector was placed 600~cm downstream from the sample in order to form propagation-based phase contrast generated by the mostly grain structure of the model kidney stone. The detector consisted of a sCMOS camera (pco.edge 5.5, PCO, Germany) optically coupled with a 1X lens tandem microscope (Hasselblad, Sweeden) \emph{via} a mirror with 500~$\mu$m LuAG:Ce (Lu$^3$Al$^5$O$^{12}$:Ce, Crytur, Che) scintillator, yielding a pixel size of 6.5~$\mu$m/pixel. During acquisition, 2000 projections of the transmitted beam were collected over a 360~$\degree$ rotation. 
Prior to the volume reconstruction, each of the recorded projections was treated with the single-distance phase retrieval method \cite{paganin2002simultaneous} utilizing the dimensionless ratio of the complex refractive index ($\delta/\beta= 300$). The volume was reconstructed with the gold-standard filtered back projection approach using the ESRF reconstruction pipeline NABU \cite{payno2022overcoming} resulting in a final volume of 2560x2560x2160 voxels. The volume was post-processed and filtered with a set of in-house developed scripts (\emph{e.g.}, ring correction as per ref.~\cite{raven1998numerical}).

\nocite{*}

\bibliography{biblio}

\end{document}